\documentclass[lettersize,journal]{IEEEtran} 

\newcommand{\sysname}{\textsc{ConneX}\xspace}
\newcommand{\artifactUrl}{\url{https://anonymous.4open.science/r/Connex-anonymous}}

\newcommand{\stitle}[1]{%
\ifhmode
    \par
\fi
\noindent\textbf{#1}%
}

\newcommand{\Yes}{\ding{52}}
\newcommand{\No}{\ding{55}}
\newcommand{\charBlock}{\rule{1em}{2ex}} 

\newcommand{\varname}[1]{\texttt{#1}\xspace} 
\newcommand{\algorithmname}{\algorithmcfname} 
\newcommand{\sectionname}{\S}
\newcommand{\linename}{Line\xspace}
\newcommand{\eqname}{Eq\xspace}

\newcommand{\alignleft}{\raggedright\arraybackslash}
\newcommand{\aligncenter}{\centering\arraybackslash}

\newcommand{\mycoloumnwidth}[1]{\hsize=#1\hsize\linewidth=\hsize}

\def\TemplateType{IEEE}

\newcommand{\TemplateACM}{\equal{\TemplateType}{acm}}
\newcommand{\TemplateIEEE}{\equal{\TemplateType}{IEEE}}
\newcommand{\TemplateUsenix}{\equal{\TemplateType}{Usenix}}
\newcommand{\TemplateIEEEJournal}{\equal{\TemplateType}{IEEE}}

\usepackage{ifthen}

\ifthenelse{\TemplateIEEE\or\TemplateUsenix}{

\usepackage{amsmath,amsfonts,bm,float,amssymb} 

}{}

\everypar{\looseness=-1}

\usepackage{amsthm}
\usepackage{sansmath} 
\usepackage{xspace}
\usepackage{graphicx} 

\usepackage{xurl} 
\usepackage{hyperref}
\hypersetup{
    colorlinks=true,
    linkcolor=blue,
    filecolor=blue,      
    urlcolor=blue,
    citecolor=blue,
    }
\usepackage[nameinlink]{cleveref}
\Crefformat{figure}{#2Figure~#1#3}
\Crefformat{table}{#2\tablename~#1#3}

\usepackage{multirow}
\usepackage{makecell}
\usepackage{caption} 
\usepackage{listings}
\usepackage[frozencache,cachedir=minted-cache]{minted}

\usepackage{setspace}  
\usepackage[linesnumbered,ruled,vlined,noend]{algorithm2e} 
\usepackage{upquote}
\usepackage{pifont}

\usepackage{array} 
\usepackage{subcaption} 

\usepackage{forest}
\usepackage{indentfirst} 

\usepackage{xcolor}

\usepackage{tikz}
\usepackage{pgfbaseimage}
\usetikzlibrary{tikzmark}
\usetikzlibrary{arrows}
\usetikzlibrary{arrows.meta}
\usetikzlibrary{patterns}
\usepackage{pgfplots}
\usepgfplotslibrary{groupplots}
\usepgfplotslibrary{statistics}
\usepackage{orcidlink}

\usepackage{threeparttable}
\usepackage{booktabs} 
\usepackage{tabularx} 
\usepackage{diagbox}

\usepackage[export]{adjustbox} 

\usepackage{wasysym} 

\usepackage{changepage}  

\usepackage{enumitem}    

\newenvironment{rqitem}{
    \begin{enumerate}[label=\textbf{RQ\arabic*}:, left=0pt, partopsep=0pt, itemsep=0pt]
}{
    \end{enumerate}
}

\usepackage{etoolbox}

\usepackage[most]{tcolorbox}
\newtcolorbox{conclusionbox}[1][]{ 
    enhanced,
	fonttitle = \large,
	colbacktitle =black,
	coltitle=white,
	attach boxed title to top left = {yshift = -5pt},
	fontupper = \small,
	colback = white,
	colframe=black,
	toprule  = -1pt,
	rightrule = -1pt,
	top = 10pt,
	title=#1
} 

\theoremstyle{plain}
\theoremstyle{definition}

\definecolor{color_of_plausible_answer}{RGB}{204,204,204}
\definecolor{color_of_plausible_answer2}{RGB}{255, 153, 153}
\definecolor{color_of_plausible_answer3}{RGB}{255, 230, 153}
\definecolor{color_of_correct_answer}{RGB}{0,204,0}
\definecolor{dark_red}{RGB}{102,0,51}
\definecolor{light_blue}{RGB}{101, 129, 178}

\newcommand{\Call}[2]{\text{#1}(#2)} 
\newcommand{\KwLength}[1]{\text{length}(#1)} 
\newcommand{\KwGet}[2]{#1.\text{get}(#2)} 
\newcommand{\KwRemove}[2]{#1.\text{remove}(#2)} 
\newcommand{\KwAdd}[2]{#1.\text{ADD}(#2)} 

\ifthenelse{\TemplateACM}{

}{\ifthenelse{\TemplateIEEE}{

\pagestyle{plain}
\ifthenelse{\TemplateIEEEJournal}{
\hyphenation{op-tical net-works semi-conduc-tor IEEE-Xplore}
\def\BibTeX{{\rm B\kern-.05em{\sc i\kern-.025em b}\kern-.08em
T\kern-.1667em\lower.7ex\hbox{E}\kern-.125emX}}

}{
\hyphenation{op-tical net-works semi-conduc-tor}
}

}{

\ifthenelse{\TemplateUsenix}{

\usepackage{templates/usenix/usenix}
\usepackage{filecontents}

}{
}
}} 

\begin{document}

\ifthenelse{\TemplateUsenix}{
    \title{\Large \bf \sysname: Automatically Resolving Transaction Opacity of Cross-Chain Bridges for Security Analysis}
}{
    \title{\sysname: Automatically Resolving Transaction Opacity of Cross-Chain Bridges for Security Analysis}
}

\ifthenelse{\TemplateACM}{





}{\ifthenelse{\TemplateIEEE}{

\author{
    Hanzhong Liang$^{\orcidlink{0009-0001-1915-6692}}$, \textit{Member, IEEE}, 
    Yue Duan$^{\orcidlink{0000-0003-1049-9645}}$, \textit{Member, IEEE},
    Xing Su$^{\orcidlink{0009-0007-5247-1444}}$, \textit{Member, IEEE},
    Xiao Li$^{\orcidlink{0009-0003-5678-4617}}$, \textit{Member, IEEE},
    Yating Liu$^{\orcidlink{0009-0007-8120-488X}}$, \textit{Member, IEEE},
    Yulong Tian$^{\orcidlink{0000-0003-0163-4542}}$, \textit{Member, IEEE},
    Fengyuan Xu$^{\orcidlink{0000-0003-3388-7544}}$, \textit{Member, IEEE},
    Sheng Zhong$^{\orcidlink{0000-0002-6581-8730}}$, \textit{Fellow, IEEE}

\IEEEcompsocitemizethanks{\IEEEcompsocthanksitem H. Liang, X. Su, X. Li, Y. Liu, F. Xu and S. Zhong are with the National Key Lab for Novel Software Technology, Nanjing University, Nanjing, China, 210023. (E-mail:
\{hanz\_liang, xingsu, xiao.li, yatingliu\}@smail.nju.edu.cn, \{fengyuan.xu,zhongsheng\}@nju.edu.cn)%
\IEEEcompsocthanksitem Y. Duan is with the Singapore Management University, Singapore. (E-mail: yueduan@smu.edu.sg)%
\IEEEcompsocthanksitem Y. Tian is with the Nanjing University of Aeronautics and Astronautics, Nanjing, China. (E-mail: yulong.tian@nuaa.edu.cn)%
\IEEEcompsocthanksitem Fengyuan Xu is the corresponding author.
}
}

}{\ifthenelse{\TemplateUsenix}{

    \author{
        {\rm Anonymous Author(s)}
} 
}{}
} 
} 

\ifthenelse{\TemplateACM}{
    
\begin{abstract}
As the Web3 ecosystem evolves toward a multi-chain architecture, cross-chain bridges have become critical infrastructure for enabling interoperability between diverse blockchain networks. However, while connecting isolated blockchains, the lack of cross-chain transaction pairing records introduces significant challenges for security analysis like cross-chain fund tracing, advanced vulnerability detection, and transaction graph-based analysis. To address this gap, we introduce \sysname, an automated and general-purpose system designed to accurately identify corresponding transaction pairs across both ends of cross-chain bridges. Our system leverages Large Language Models (LLMs) to efficiently prune the semantic search space by identifying semantically plausible key information candidates within complex transaction records. Further, it deploys a novel examiner module that refines these candidates by validating them against transaction values, effectively addressing semantic ambiguities and identifying the correct semantics. Extensive evaluations on a dataset of $\sim$500,000 transactions from five major bridge platforms demonstrate that \sysname achieves an average F1 score of 0.9746, surpassing baselines by at least 20.05\%, with good efficiency that reduces the semantic search space by several orders of magnitude (1e10 to less than 100). Moreover, its successful application in tracing illicit funds (including a cross-chain transfer worth \$1 million) in real-world hacking incidents underscores its practical utility for enhancing cross-chain security and transparency.
\end{abstract}





\newcommand{\ynote}[1]{\textcolor{blue}{\bf \emph{yulong: #1}}}

\maketitle

}{\ifthenelse{\TemplateIEEE}{

\maketitle

\begin{abstract}
As the Web3 ecosystem evolves toward a multi-chain architecture, cross-chain bridges have become critical infrastructure for enabling interoperability between diverse blockchain networks. However, while connecting isolated blockchains, the lack of cross-chain transaction pairing records introduces significant challenges for security analysis like cross-chain fund tracing, advanced vulnerability detection, and transaction graph-based analysis. To address this gap, we introduce \sysname, an automated and general-purpose system designed to accurately identify corresponding transaction pairs across both ends of cross-chain bridges. Our system leverages Large Language Models (LLMs) to efficiently prune the semantic search space by identifying semantically plausible key information candidates within complex transaction records. Further, it deploys a novel examiner module that refines these candidates by validating them against transaction values, effectively addressing semantic ambiguities and identifying the correct semantics. Extensive evaluations on a dataset of $\sim$500,000 transactions from five major bridge platforms demonstrate that \sysname achieves an average F1 score of 0.9746, surpassing baselines by at least 20.05\%, with good efficiency that reduces the semantic search space by several orders of magnitude (1e10 to less than 100). Moreover, its successful application in tracing illicit funds (including a cross-chain transfer worth \$1 million) in real-world hacking incidents underscores its practical utility for enhancing cross-chain security and transparency.
\end{abstract}



}{\ifthenelse{\TemplateUsenix}{
    \maketitle

\begin{abstract}
As the Web3 ecosystem evolves toward a multi-chain architecture, cross-chain bridges have become critical infrastructure for enabling interoperability between diverse blockchain networks. However, while connecting isolated blockchains, the lack of cross-chain transaction pairing records introduces significant challenges for security analysis like cross-chain fund tracing, advanced vulnerability detection, and transaction graph-based analysis. To address this gap, we introduce \sysname, an automated and general-purpose system designed to accurately identify corresponding transaction pairs across both ends of cross-chain bridges. Our system leverages Large Language Models (LLMs) to efficiently prune the semantic search space by identifying semantically plausible key information candidates within complex transaction records. Further, it deploys a novel examiner module that refines these candidates by validating them against transaction values, effectively addressing semantic ambiguities and identifying the correct semantics. Extensive evaluations on a dataset of $\sim$500,000 transactions from five major bridge platforms demonstrate that \sysname achieves an average F1 score of 0.9746, surpassing baselines by at least 20.05\%, with good efficiency that reduces the semantic search space by several orders of magnitude (1e10 to less than 100). Moreover, its successful application in tracing illicit funds (including a cross-chain transfer worth \$1 million) in real-world hacking incidents underscores its practical utility for enhancing cross-chain security and transparency.
\end{abstract}
}{}

}}


\section{Introduction}



The Web3 ecosystem is transitioning towards a multi-chain architecture, with decentralized applications (DApps) operating across a number of layer-1s (e.g., Ethereum~\cite{Ethereum_yellow_paper}, Solana~\cite{Solana}), layer-2 solutions (e.g., Optimism~\cite{Optimism_doc}, Arbitrum~\cite{Arbitrum}), and app-specific chains (e.g., Avalanche subnets~\cite{Avalanche_doc}). This paradigm allows developers to leverage the distinct advantages of different networks, fostering innovation and diversity. However, a fundamental limitation of blockchains is their design as isolated systems, incapable of natively communicating with external entities, including other blockchains. This inherent lack of direct communication creates a significant barrier to interoperability~\cite{augusto2024sok}. Consequently, cross-chain bridge platforms (e.g., Stargate~\cite{Stargate}) have emerged as a critical solution. A bridge acts as a connector between any pair of blockchains, enabling the transfer of assets, data, or functionalities across networks that otherwise operate independently. 

By linking disparate chains, cross-chain bridges play a vital role in unlocking the full potential of the multi-chain Web3 ecosystem. For instance, they allow users to move tokens or other digital assets from one blockchain to another seamlessly. As a result, cross-chain bridges are widely used by DeFi aggregators (e.g., 1inch~\cite{1inch}) or lending protocols (e.g., Compound~\cite{Compound}), with over \$1 billion locked onchain~\cite{DefiLlama}. 
A cross-chain transfer procedure usually consists of a transaction on the source chain, a transaction on the destination chain, and an off-chain monitoring component (i.e., the bridge). 
The correspondence relationship between the source transaction and the destination chain transaction is termed \textit{cross-chain transaction pair}.

\begin{figure}[!t]
    \centering
    \includegraphics[width=0.88\linewidth]{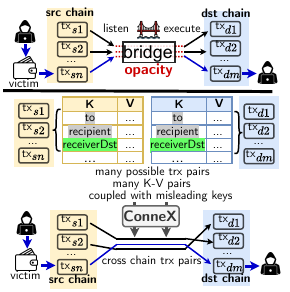}
    \caption{The cross-chain transaction opacity problem and our solution, \sysname. \textbf{Top:} Cross-chain bridges introduce \textit{transaction opacity} by breaking the direct relationship between source and destination transactions, a vulnerability that attackers exploit to launder funds. \textbf{Bottom:} Re-establishing this pairing relationship is challenging due to the massive data volume and the difficulty of distinguishing authentic semantic keys from numerous possibilities and those misleading ones within transaction data. By overcoming these challenges, \sysname reconstructs the pairings, enabling crucial security analyses, such as tracing illicit funds (blue arrow) across different blockchains.}
    \label{fig:our_work}
\end{figure}



\stitle{Security Impacts.} Despite their utility for interoperability, cross-chain bridges introduce significant challenges for security analysis by obfuscating transactional data flows and create transaction opacity between blockchains. First, cross-chain bridges complicate or even impede the process of tracing funds across blockchains, which is a critical aspect of security-related analysis. For example, studies investigating money laundering activities~\cite{ccn2025crypto-criminals,chananalysis2024MoneyLaundering,wu2023toward} show that they often trace illicit funds being transferred through cross-chain bridges, which terminate further investigation due to the absence of the corresponding pairs~\cite{wu2023toward}. Second, a comprehensive understanding of code semantics, essential for advanced vulnerability detection is unachievable without cross-chain transaction records~\cite{10.1145/3678890.3678894}.
Finally, the data opacity caused by bridges constrains other key analyses, including graph-based behavioral modeling~\cite{zhou2022behavior} and the study of cross-chain Miner Extractable Value (MEV) and sandwich attack\cite{ferreira2024rolling}.
In summary, resolving the opacity for cross-chain bridge platforms and integrating third-party participation are paramount to ensuring the long-term robustness and security of the Web3 ecosystem. Consequently, there is a pressing need for an automated tool capable of accurately identifying cross-chain transaction pairs to lay a crucial foundation for security research.

However, establishing accurate cross-chain relationships poses significant challenges. First, some bridges prioritize privacy or security~\cite{Portal, zkBridge}, which results in opaque or inaccessible transaction records. 
Second, researchers have highlighted the inherent complexity of cross-chain logic~\cite{belchior2023hephaestus}, emphasizing the diverse and decentralized nature of bridge implementations. Additionally, the vast volume of transaction data, combined with intricate and interconnected relationships across multiple chains, further complicates the accuracy of matching cross-chain transactions.

\stitle{Insight.} Our insight is that while the on-chain records of a single cross-chain transaction may differ structurally (i.e., comprising different elements within their respective collections) between the source and destination chains, the underlying semantic intent of the transactions must correspond~\cite{zhang2022xscope, eshghie2024highguard}. 
This semantic correspondence serves as an identifier that distinguishes a cross-chain transaction pair from other pairs. Specifically, the intent and effect of a cross-chain transaction, namely the transfer of a specific asset from one chain to another, must be consistently reflected on both sides of the bridge. To capture the intent of the transactions from each side, we define a `semantic quintuple' comprising destination, chain, amount, asset type, and timestamp (detailed in \sectionname~\ref{subsec:symbols}). This quintuple is derived from the transaction records, and the combination of the two quintuples from two ends can characterize the underlying semantics of a cross-chain transaction. 
By identifying and comparing these semantic quintuples for transactions on both ends, it becomes feasible to accurately identify corresponding cross-chain transaction pairs by analyzing publicly available on-chain data.






\stitle{Technical Challenges.} However, extracting accurate identifiers from any transaction presents two major technical challenges. 
The first challenge is the semantic searching space explosion (\textbf{C1}). Transactions and their associated event logs constitute complex, often deeply nested, data structures. A single transaction may contain dozens to hundreds of fields\footnote{In scenarios where multiple structured data are processed collectively (e.g., a transaction and its event logs), key collisions can occur as the same key name may appear. To resolve this ambiguity, these related structures are conceptualized as a forest. A \texttt{field} is a unique identifier created by concatenating the semantic names of all nodes along the path from a root to a leaf node (\sectionname~\ref{subsec:preprocessing}).}. When it comes to the possible combination of choosing five keys from such a collection, the result easily reaches millions (e.g., $C_{100}^{5}$). When scaled by millions of daily transactions across multiple blockchains~\cite{DefiLlama}, each with a potential heterogeneous data structure, this search becomes computationally intractable. Consequently, a brute-force enumeration approach to identifying quintuples becomes computationally infeasible due to the sheer scale. 
The second challenge lies in the ambiguous or sometimes misleading semantics (\textbf{C2}). Multiple fields may exhibit similar meanings, making it extremely difficult even for experienced human analysts to distinguish the correct correspondence. Furthermore, the ground-truth semantic elements are not always the most intuitively obvious (e.g., ground truth `receiverDst' v.s. misleading `target'), meaning that simple similarity-based selection methods will be error-prone. To the best of our knowledge, no technique exists so far that can automatically and efficiently extract the identifier quintuple from such a vast and complex semantic search space.

\stitle{Our Solution.} To address these challenges, we propose \sysname, an automated system to accurately determine cross-chain transaction pairs, by employing a novel semantic-aware key-value dual pruning method, which efficiently identifies corresponding quintuples within the vast search space of possible combinations.
As shown in \figurename~\ref{fig:our_work}, \sysname ingests transaction records from both ends of a given bridge and outputs a set of identified transaction pairs. These pairs serve as a foundational data source for a wide range of downstream security tasks, such as anti-money laundering or cross-chain attack analysis. 
The core objective of \sysname is to identify the correct semantic quintuple from a vast, developer-defined name space (the key collection), which is then used for generating pairs. 
To achieve this, \sysname first preprocesses raw input transactions into a structured key-value pair format (\sectionname~\ref{subsec:preprocessing}) and employs a preliminary classification step, which groups transactions that share identical key collections (\sectionname~\ref{subsec:categorizing}), thereby reducing the overall number of candidate quintuples requiring individual processing.
Following this, \sysname implements a two-step pruning procedure. It incorporates a semantic filtering step (acting on a collection of keys) and a value-based semantic filtering step (utilizing `values' for further key pruning).
To address \textbf{C1}, \sysname leverages an LLM that performs semantic filtering on the entire key space, rapidly identifying a reduced set of plausible key candidates for each element of the quintuple (\sectionname~\ref{subsec:query_llm}). Then, to address \textbf{C2}, we incorporate a novel examiner module (\sectionname~\ref{subsec:validation}). This examiner acts as a verification mechanism, refining the LLM's proposed candidate quintuples to the truly valid ones and filtering out potentially misleading candidates. The verification introduces the `values' component of the key-value pair as a new factor to prune the ambiguous candidate quintuples. 
Specifically, the examiner extracts critical semantic values such as participants' addresses, their associated token inflows/outflows, and cross-chain destinations within a given transaction. This extracted information constitutes the `values' used to validate the candidate semantic quintuples (the `keys'). The scope of candidate keys is consequently shrunk if the extracted values do not align with a valid cross-chain intent.
By combining the LLM's ability to efficiently explore the semantic searching space with the examiner's ability to validate the correct semantics, \sysname achieves both high accuracy and scalability in identifying semantic quintuples.

We implement the prototype of \sysname and conduct extensive evaluations using a dataset comprising $\sim$500,000 transactions collected from five mainstream cross-chain bridge platforms (Stargate, DLN, Multi, Celer, Poly) over the period of February 2021 to March 2024. \sysname attains an average F1 score of 0.9746, outperforming at least 20.05\% over baselines (\sectionname~\ref{subsec:RQ1}). Meanwhile, the core pruning method of\sysname reduces the search space by several orders of magnitude (from over $10^{10}$ to fewer than 100 candidates). This results in an average processing time of 0.4 seconds per transaction, a speedup of up to 9x compared to the baseline (\sectionname~\ref{subsec:abalation}). Moreover, we demonstrated the practical utility of \sysname for the downstream task of cross-chain money laundering analysis with real-world cases (\sectionname~\ref{subsec:application}). By integrating \sysname with existing fund tracing techniques~\cite{wu2023tracer,wu2023toward}, we identify transactions involving cross-chain fund transfers by hackers and successfully flag their intermediary addresses on the destination chains. Notably, in the Bybit Hack, the largest theft of funds to date, \sysname successfully identifies a transit address receiving \$1 million USDCs on Solana.


In summary, this paper makes the following contributions:

\begin{itemize}
    \item We design, to the best of our knowledge, the first \textit{automated} and \textit{generalized} discovery mechanism of hidden cross-chain transaction pairs between two blockchains, named \sysname\footnote{available at \artifactUrl}. \sysname operates without requiring any collaboration from blackboxed bridge platforms, ensuring general applicability to diverse cross-chain bridge platforms. 
    \item We define a general semantic quintuple as a robust identifier for cross-chain transactions, and propose a novel semantic-aware key-value dual pruning method to efficiently match corresponding quintuples within the vast search space of possible combinations. 
    \item We conduct extensive experiments using a substantial dataset of real-world cross-chain transactions to evaluate the performance of \sysname. The results demonstrate the effectiveness of our approach, achieving an average F1 score of 0.9746 and improving at least 20.05\% over baselines. 
    \item We showcase how \sysname can be applied to analyze and track cross-chain money laundering activities by successfully identifying instances of illicit fund transfers across blockchains in two real-world hacking incidents.
\end{itemize}



\section{Background \& Related Work}
\label{sec:background_and_motivation}


\subsection{Blockchain}
\label{subsubsec:background_blockchain}



Blockchain is a distributed ledger technology where data is stored in blocks that are cryptographically linked to form an immutable chain. Each block typically contains a set of transaction records.
Blockchains are primarily classified by their access control model, Public (permissionless) blockchain is open to any participant, e.g., Bitcoin and Ethereum~\cite{Ethereum_yellow_paper}. Private (permissioned) blockchain Access is restricted to authorized participants, e.g., Hyperledger Fabric~\cite{hyperledger_fabric_project}.

\subsection{Cross-chain Bridge and Transaction}
\label{subsubsec:background_crosschain}

\begin{figure}[!t]
    \centering
    \includegraphics[width=1.0\linewidth]{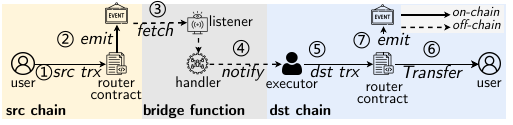}
    \caption{The Cross-Chain Transaction Model.}
    \label{fig:cross_chain_model}
\end{figure}

Cross-chain bridges are applications designed to exchange assets across different blockchain networks, enabling asset transfers from the source chains to the target chains. As an implementation of interoperability~\cite{wang2021sok}, they typically consist of three components: asset deposit on the source chain, asset withdrawal on the target chain, and an intermediary mechanism connecting the source and target chains. This end-to-end process is defined as a \textit{cross-chain transaction}~\cite{belchior2023hephaestus}. 
While the user typically initiates the source transaction, the target transaction is often triggered by an automated component of the bridge, such as a relayer or validator network.
Bridges can be categorized to \textit{public} and \textit{private} based on the availability of the cross-chain transaction records.
Bridge architectures vary. As illustrated in \figurename~\ref{fig:cross_chain_model}, a common design utilizes smart contracts on the source and destination chains, connected by off-chain relayers that monitor events on source and trigger corresponding actions on the destination. They can employ different mechanisms such as Lock-and-Mint, Burn-and-Release, or liquidity pool-based approaches~\cite{10.1145/3678890.3678894}. 
As shown in \figurename~\ref{fig:cross_chain_model}, after a user initiates a transaction on the source chain (\ding{172}), the off-chain relayer monitors and parses relevant events emitted (\ding{173}-\ding{174}), notifying the executor on the destination chain (\ding{175}). The executor then initiates a transaction on the destination chain (\ding{176}), enabling the transfer of assets or data to the target receiver (\ding{177}-\ding{178}).

\stitle{Event Logs.} Event logs are a structured logging mechanism native to the Ethereum Virtual Machine (EVM), serving as a primary interface for smart contracts to communicate runtime information to external applications~\cite{soliditydoc}. Technically, logs are encoded data that, once decoded off-chain using the contract's Application Binary Interface (ABI), provide verifiable information about contract execution. This enables diverse functionalities, from updating user interfaces to facilitating complex protocols~\cite{li2023understanding}.
In cross-chain environments, event logs often carry critical identifiers about input assets, users, or inter-chain messages~\cite{DLN_DeveloperDoc}. Consequently, any misinterpretation or mishandling of these logs can lead to state inconsistencies between the bridged chains, potentially causing significant financial losses~\cite{10.1145/3678890.3678894, lee2023sok}.

\subsection{Related Work}
\label{subsec:exsiting_work}

\stitle{Cross-Chain Communication and Interoperability.}
Existing research on cross-chain communication and interoperability~\cite{zamyatin2021sok, belchior2021survey, buterin2016chain, ou2022overview, wang2021sok, kotey2023blockchain, han2023survey, augusto2024sok} primarily focuses on identifying the challenges and opportunities in enabling seamless asset and data exchange across diverse blockchain ecosystems. In addition, some studies delve into cross-chain protocol implementation~\cite{guo2024zkcross, frauenthaler2020eth} or evaluation~\cite{kang2022blockchain, chervinski2023analyzing}. These works show the fragmented nature in cross-chain ecosystem, highlighting the necessity of a unified approach for evaluating and analyzing.

\stitle{Cross-Chain Security Surveys.}
Lee et al.~\cite{lee2023sok} decompose the cross-chain bridge design and introduce the risks associated with each bridge components. 
Haugum et al.~\cite{haugum2022security} provide the security and privacy challenges in blockchain interoperability.
Zhang et al.~\cite{10.1145/3678890.3678894} categorize both the cross-chain bridge and their attack incidents.



\stitle{Cross-Chain Traceability and Monitoring.}
Researcher propose cross-chain architectures such as zkCross~\cite{guo2024zkcross} and \cite{cao2023cross} to enhance cross-chain traceability. Hinteregger et al.~\cite{hinteregger2019short} try to trace Monero transactions.
XChainWatcher~\cite{augusto2024xchainwatcher} and HighGuard~\cite{eshghie2024highguard} propose monitoring and detecting framework based on formal specifications. The two works aim to analyze cross-chain attack and monitor business logic violations. 

\stitle{Cross-Chain Security Analysis.}
There exists a line of research that works on studying cross-chain transactions and their security or performance issues. Xscope~\cite{zhang2022xscope} is the first detection tool that focuses on existing bridge attacks. They formalize the on-chain and off-chain actions into logic representations, and use an SMT solver to detect violations. 
SmartAxe~\cite{liao2024smartaxe} further proposes a static framework to analyze the vulnerabilities in cross-chain contracts. 
Hephaestus~\cite{belchior2023hephaestus} models cross-chain transactions and builds Hyperledger Cacti components to evaluate the cross-chain performance. 
Connector~\cite{lin2025connector} and Xsema~\cite{zheng2024xsema} propose semantic extraction methods to identify whether a transaction is cross-chain related. 

These works highlight the necessity of our work in two key aspects.
First, many of these techniques (e.g., cross-chain attack analysis~\cite{zhang2022xscope} or cross-chain monitoring~\cite{belchior2023hephaestus, augusto2024xchainwatcher}) presume the availability of cross-chain transaction pairing relationships—the very output of our research—for their proper execution. While Connector~\cite{lin2025connector} offers a rule-based solution for generating these pairings, it also faces challenges of the second aspect - limited scope. 
Specifically, the rule-based design inherent in most existing approaches restricts their applicability to a limited number of bridges (e.g., Xscope on THOR, Hephaestus on self-made toy bridges, Connector on three bridges). In contrast, our generalized cross-chain model overcomes this limitation.
\textit{To the best of our knowledge, our work is the first generalized, automated tool that generates cross-chain pairing relationships.}


\section{Motivating Example}
\label{sec:motivating_example}

\begin{figure}[!t]
    \centering
    \subfloat[Challenge 1: Semantic Searching Space Explosion]{
        \label{fig:motivation_example_C1}
            \begin{tabular}{rrr}
            \toprule
            \textbf{\# logs} & \textbf{\# fields} & \textbf{\# combinations} \\
            \midrule
            9              & 144                & 481,008,528                      \\
            \bottomrule
            \end{tabular}

    }
    \vspace{0.3em}

    \subfloat[Challenge 2: Ambiguous or Misleading Semantics. {\color{color_of_plausible_answer}\charBlock} (misleading semantics). {\color{color_of_correct_answer}\charBlock} (correct semantics)]{
        \label{fig:motivation_example_C2}
        \includegraphics[width=0.8\linewidth]{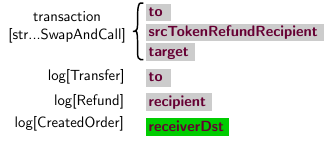}
    }
    \caption{Technical Challenges}
    \label{fig:motivation_example}
\end{figure}

\figurename~\ref{fig:motivation_example} illustrates the challenges encountered when extracting quintuples(defined in \sectionname~\ref{subsec:symbols}) from a real-world transaction\footnote{\url{https://etherscan.io/tx/0xbbb00ccff6a1794a6f4bc6b3eb46119db25e5a31d77bf24d9f1a22ea3b5751b1}}, which is then used for generate final pairings. These challenges primarily include: (\ref{fig:motivation_example_C1}) combinatorial search space explosion and (\ref{fig:motivation_example_C2}) ambiguous or misleading semantics.
First, the search space for relevant data fields is immense(\textbf{C1}). The example transaction contains 9 event logs that yield 144 distinct fields. A brute-force approach to identify a 5-element quintuple from this set would face a computationally infeasible number of possible combinations ($C_{144}^{5}$=481,008,528). This staggering number represents the search space for just \textit{a single transaction}. When considering heterogeneous transactions (i.e., transactions with varying fields) involved in cross-chain operations, the search space expands to an unmanageably vast scale.
Second, transactions exhibit significant semantic ambiguity(\textbf{C2}). Multiple fields may appear synonymous but are contextually distinct. For instance, the example contains several fields that could represent a `destination`, such as a `recipient` field within a `Refund` event. However, only one of these fields (highlighted in {\color{color_of_correct_answer}\charBlock}) corresponds to the true target address of the cross-chain transfer. Incorrectly selecting other misleading candidates (highlighted in {\color{color_of_plausible_answer}\charBlock}) leads to erroneous data extraction, resulting in False Positives (FP) or False Negatives (FN) when generating cross-chain transaction pairings.

To address these challenges, \sysname incorporates two core components. 
First, to effectively filter from multiple semantic candidates, \sysname leverages an LLM to select the candidate field that most closely aligns with the intended semantic.
Second, to mitigate the effect of misleading information, \sysname employs an examiner module to validate the correctness of candidate fields and prune the search space to one. 
This design is detailed further in \sectionname~\ref{sec:design}. 

\section{Problem Statement}

\label{sec:problem_definition}

This section establishes the formal groundwork for our analysis. We define the core terminology and symbols (\sectionname~\ref{subsec:symbols}), outline the problem scope (\sectionname~\ref{subsubsec:scope_and_assumption}) and introduce our generalized cross-chain model (\sectionname~\ref{subsec:cross_chain_model}). 


\subsection{Symbol Definitions} 

\label{subsec:symbols}


\stitle{Fields.} To prevent key collisions when processing multiple nested data structures (e.g., a transaction and its associated event logs), we define a unique identifier for each data point. A \texttt{field} is a unique path identifier, constructed by concatenating the semantic names of nodes from a root to a leaf in the data structure (examples in \sectionname~\ref{subsec:preprocessing}).

\stitle{Transactions and Event Logs.} A blockchain transaction, $t$, is an atomic data unit. An event, $e$, is a log emitted during a transaction's execution. We model both as structured key-value mappings from a set of fields ($F$) to corresponding values ($V$):

\begin{equation}
    (F_1, V_1), ..., (F_n, V_n) \label{eq:fields}
\end{equation}

Here, fields ($F_i$) are static, human-readable identifiers, while values ($V_i$) are dynamic, instance-specific data. These mappings can be nested, meaning that a value can be another mapping, allowing for complex data structures.

\stitle{Transactions Instance.} A \textit{transaction instance}, $tx$, comprises a single transaction $t$ and all associated event logs. The set of all transaction instances is denoted by $TX$.


\stitle{Quintuple.} Cross-chain transaction draws a direct parallel to the fundamental requirements of real-world international financial transfers. In such traditional cross-border payments, unique identification hinges on specifying the recipient's account, the corresponding country/bank, the currency type, the exact amount, and the transaction timestamp. Analogously, a quintuple includes 5 pieces of key information, namely:

\begin{itemize}[itemsep=0pt]
    \item $K_D$: The destination address (i.e., recipient's wallet address)
    \item $K_C$: The counterpart blockchain (defined as `source chain' when considering a transaction on destination chain, and vice versa.)
    \item $K_T$: Asset types used by the sender and received by the recipient (note these may differ, e.g., ETH sent, WETH received)
    \item $K_A$: Total amount of assets expended by the sender and received by the recipient (note the latter is usually less due to bridge operation fees)
    \item $K_{Ts}$: Timestamp of the cross-chain transaction on both ends
\end{itemize}

Formally, a quintuple is a mapping from these semantic keys to their corresponding field identifiers within a transaction instance: $\{K_i \mapsto F_i\}$. It serves as a template to locate the specific data points ($K_D, K_C, \dots$) within any given $tx$.

\subsection{Problem Definition and Scope}
\label{subsubsec:scope_and_assumption}


\stitle{Our goal.} 
The primary objective of this work is to develop a method for systematically identifying and pairing cross-chain transactions. Given transaction instances from a source chain ($TX_s$) and a destination chain ($TX_d$), we aim to find all pairs $(tx_s, tx_d)$, where $tx_s \in TX_s$ and $tx_d \in TX_d$, such that $tx_d$ is the direct result of $tx_s$ being processed by a bridge. We denote this relationship as $Bridge(tx_s) = tx_d$. This model will be applied across all blockchain pairs supported by a given bridge platform.

\stitle{Blockchain Scope.} Our analysis is limited to public, EVM-compatible blockchains. This is justified as they represent the majority of top-ranked chains (\tablename~\ref{table:chains_rank}). Non-EVM chains (e.g., Bitcoin) and private networks are out of scope.

\stitle{Bridge Scope.} We focus on third-party, contract-based bridges where a cross-chain transfer consists of a single transaction on the source chain and a corresponding single transaction on the destination chain(also named \textit{final transaction} in \cite{augusto2024xchainwatcher}). This model is representative of most popular bridges (\tablename~\ref{table:bridges_rank}). Consequently, we exclude: (1) Bridges that do not operate via smart contracts (e.g., Avalanche Bridge~\cite{Avalanche_doc}). (2) Canonical bridges that are an integral part of a Layer 2's native protocol, as they employ unique security and verification models beyond our generalized approach.


\begin{table}[!t]
	\centering        
    \footnotesize
    \caption{The Top 10 Chains Ranked by Total Value Locked. Retrieved from Chainspot~\cite{chainspot} in May 2024.}
    \label{table:chains_rank} 
    \begin{tabularx} 
        {.75\linewidth}{l|>{\aligncenter}X>{\aligncenter}X>{\aligncenter}X} 
    \toprule
    \textbf{name}  &  \textbf{public?}  & \textbf{event logs?} & \textbf{VM-support} \\ 
    \midrule
    Ethereum  ~\cite{Ethereum_yellow_paper}& \Yes  & \Yes & EVM \\
    Solana    ~\cite{Solana}               & \Yes  & \Yes$^{\flat}$ & SVM \\
    BSC       ~\cite{BSC_doc}              & \Yes  & \Yes & EVM \\
    Arbitrum  ~\cite{Arbitrum}             & \Yes  & \Yes & EVM \\
    Base      ~\cite{Base_doc}             & \Yes  & \Yes & EVM \\
    THORchain ~\cite{THORchain_doc}        & \Yes  & \No & - \\
    Optimism  ~\cite{Optimism_doc}         & \Yes  & \Yes & EVM \\
    Sui       ~\cite{Sui_doc}              & \Yes  & \Yes & Move \\
    Avalanche ~\cite{Avalanche_doc}        & \Yes  & \Yes & Coreth$^{\natural}$ \\
    zkSync Lite ~\cite{zkSync_doc}         & \Yes  & \Yes & zkEVM$^{\natural}$ \\
    \bottomrule 
    \end{tabularx}
    \caption*{$^{\flat}$: Unstructured log data. $^{\natural}$: EVM-compatible.}
\end{table}

\begin{table}[!t]
	\centering        
    \footnotesize
    \caption{The Top 10 Cross-Chain Bridges Ranked by Total Value Locked. Retrieved from Chainspot~\cite{chainspot} in May 2024. }
    \label{table:bridges_rank} 
    \begin{tabularx} 
        {0.8\linewidth}{
            >{\mycoloumnwidth{1.9}\alignleft}X|
            >{\mycoloumnwidth{0.7}\aligncenter}X
            >{\mycoloumnwidth{0.7}\aligncenter}X
            >{\mycoloumnwidth{0.7}\aligncenter}X} 
    \toprule
    \textbf{name}    & \textbf{public?} & \textbf{contract?} & \textbf{compatible?} \\ 
    \midrule
    Across~\cite{Across_bridge}     & \No & \Yes & \Yes \\
    Arbitrum Bridge~\cite{Arbitrum_bridge}  & \No  & \No  & \Yes \\
    zkBridge~\cite{zkBridge}        & \No  & \Yes &  $\mathcal{?}^{\flat}$ \\
    DLN~\cite{DLN_bridge}  & \Yes & \Yes & \Yes \\
    Stargate~\cite{Stargate}    & \Yes$^{\natural}$  & \Yes & \Yes \\
    Meson~\cite{Meson}          & \Yes & \Yes & \No \\
    Celer~\cite{Celer_cBridge}     & \No  & \Yes & \Yes \\
    Portal~\cite{Portal}  & \No  & \Yes & \Yes \\
    Synapse~\cite{Synapse}  & \Yes & \Yes & \Yes \\
    Base Bridge~\cite{Base_doc}       & \No  & \No & $\mathcal{?}^{\flat}$ \\
    \bottomrule 
    \end{tabularx}
    \caption*{$^{\natural}$:Third-party records. \newline $^{\flat}$: Unknown due to lack of description in document.}
\end{table}



\subsection{Cross-Chain Model}
\label{subsec:cross_chain_model}




\figurename~\ref{fig:cross_chain_model} illustrates our generic cross-chain model. The workflow of a cross-chain transaction is as follows: \ding{172} A user initiates a transaction on the source chain by calling a bridge's router contract. \ding{173} Upon execution, this contract processes the transfer and emits event logs containing its details. \ding{174} An off-chain component, which we abstract as a black-box function $Bridge(\cdot)$, monitors these events. This component (e.g., a relayer network or an automated service) is responsible for propagating the transaction information to the destination chain. \ding{175} Triggered by the off-chain relay or user, \ding{176} a transaction is executed on the destination chain. This transaction interacts with the destination router contract to \ding{177} complete the asset transfer (e.g., by minting equivalent tokens) and \ding{178} emits a final set of event logs confirming completion.


This model is designed abstract to ensure generalization, distinguishing it from prior work~\cite{zhang2022xscope, belchior2023hephaestus, 10.1145/3678890.3678894}. First, as opposed to existing research (e.g., \cite{zhang2022xscope}), it is agnostic to event types, imposing no restrictions on the structure or signature of events emitted by the bridge contracts. Second, it abstracts the off-chain logic ($Bridge(\cdot)$), eliminating the requirements of its specific details. Third, it is independent of the security mechanism (unlike \cite{10.1145/3678890.3678894}), remaining compatible with various cross-chain communication protocols. Our analysis (see \tablename~\ref{table:bridges_rank}) reveals that, among the top 10 bridges ranked by Total Value Locked (TVL), 7 are compatible with our model.



\section{System Design}
\label{sec:design}

\subsection{Design Overview}
\label{subsec:design_overview}

\begin{figure*}[t]
    \centering
    \includegraphics[width=0.95\linewidth]{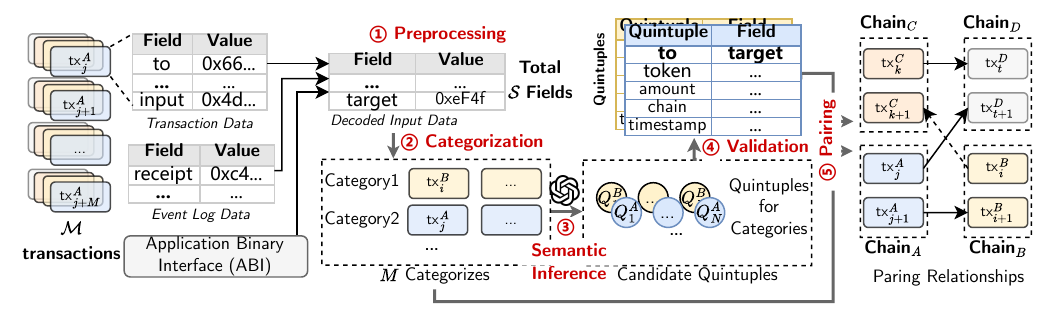}
    \caption{The Workflow of \sysname.}
    \label{fig:system_workflow}
\end{figure*}

\sysname processes all cross-chain related transactions and their associated event logs from two blockchain networks along with the corresponding ABIs and outputs the pairing relationships found within these records. The procedure of \sysname can be conceptualized as a pruning mechanism, designed to distill hundreds of thousands of potential semantic interpretations into a manageable set which is then used for cross-chain pairing. As depicted in Figure~\ref{fig:system_workflow}, \sysname consists of five major steps.
\ding{172} Preprocessing: given the raw input data and ABI specification, \sysname decodes each transaction and event logs into mapping format, representing fields and their corresponding values (\sectionname~\ref{subsec:preprocessing}).
\ding{173} Categorization: then, \sysname performs a step to categorize the decoded transactions, which involves grouping semantically identical transactions based on predefined semantics (i.e., fields) provided by bridge developers(\sectionname~\ref{subsec:categorizing}). 
\ding{174} Semantic Inferring: for each data category, \sysname further queries an LLM to rapidly filters the extensive candidate space and obtain a set of candidate semantics pertaining to the identifier quintuple (\sectionname~\ref{subsec:query_llm}). 
\ding{175} Validation: to identify the correct quintuple from these candidates, \sysname employs an examination algorithm that further prunes the possible combination to a valid one (\sectionname~\ref{subsec:validation}). 
\ding{176} Pairing: using the validated key information, \sysname attempts to match against all decoded transactions and outputs the resulting pairing relations (\sectionname~\ref{subsec:pairing}).


\subsection{Preprocesssing}
\label{subsec:preprocessing}

To enable semantic analysis, \sysname decodes raw blockchain transaction data into a structured, semantically rich format. This process takes two inputs: the raw transaction, which includes user-provided hexadecimal input data, and the corresponding smart contract's ABI.
The ABI serves as a schema, defining the static structure of function calls, including parameter names, types, and encoding rules. By applying the ABI to the transaction's input data, \sysname generates decoded name-value pairs (as in \eqname~\ref{eq:fields}). In this output, the parameter names are derived from the static ABI, while the values are determined dynamically at runtime from the user's input.

The decoded data is structured and represented in a nested JSON format to represent hierarchical relationships. This structure can be conceptually represented as a multi-way tree, where leaf nodes represent the values of the data, and the non-leaf nodes along the path from the root to a leaf represent the keys used to access those values. We define a \texttt{field} as the ordered concatenation of all keys associated with a particular value. The purpose of a field is to capture the static semantics within a transaction instance, preserving the semantic information from the root to the leaf. For instance, the field \varname{transaction[strictlySwapandCall].target} indicates the \textit{target} parameter within the \varname{strictlySwapandCall} function of the transaction.

\subsection{Categorization}
\label{subsec:categorizing}

The categorization step aims to categorize transaction instances into semantically equivalent groups based on their fields. This categorization process is unbounded, allowing for the identification of new transaction types as they emerge. To achieve this, \sysname takes preprocessed transaction data as input, performs categorization, and outputs the data organized by category. 
Specifically, a transaction category $C_S$ is defined by a unique set of fields. All transactions $tx$ within a given category share the exact same set of fields, denoted as $\mathcal{F}(tx) = \{F_1, F_2, \dots, F_n\}$.
Therefore, for any $tx_a$ and $tx_b$, they belong to the same class if and only if $\mathcal{F}(tx_a) = \mathcal{F}(tx_b)$. 
Each category represents different functions invoked by user transactions or distinct handling path taken by a cross-chain bridge platform when processing user requests. 

For example, the Stargate bridge exhibits different fields depending on the requested asset: (1) If a user requests ETH as the output asset, but the bridge doesn't directly support ETH withdrawals, it will transfer a self-issued wrapped ETH token (SGETH) to the user instead, resulting in a field like \varname{log[TransferNative]...}; (2) Otherwise, if the requested asset (typically an ERC-20 token) is supported, it will directly transfer the asset to the user-specified address, resulting in a field \varname{log[Transfer]...}.

After grouping a set of transactions into $M$ distinct categories, where each category $i$ is defined by a field set $S_i$, we calculate the total number of possible quintuple selections, $\mathcal{X}$, across all categories:

\begin{equation}
    \mathcal{X}=\sum_{i=1}^{M} C_{|S_i|}^{5}  \label{eq:possible_combination1}
\end{equation}

which represents the sum of combinations of choosing 5 fields from the $S_i$ within each category.

\subsection{Semantic Inference}
\label{subsec:query_llm}



This step performs semantic inference by harnessing the power of LLM, with the goal of identifying candidate fields that exhibiting semantic similarity to a quintuple (defined in \sectionname~\ref{subsec:symbols}) from a vast space. 
\sysname ingests the categorized data from \sectionname~\ref{subsec:categorizing} and outputs possible quintuple candidates $QT_{cdd}^{i}$ for each category $i$ defined by a field set $S_i$. 

The process involves sampling $N$ transaction instances from each category to use as contextual examples for the LLM. We employ a sophisticated prompt engineering strategy that combines several techniques to improve accuracy: (1) Domain Scoping: The LLM's reasoning is constrained to the blockchain and cross-chain domains. (2) Few-Shot Learning: A manually curated example, including the corresponding answer and a chain-of-thought explanation~\cite{wei2022chain}, is included in the prompt to guide the model. (3) Structured Output: The prompt requires the LLM to return its findings in a JSON format, including a confidence score for each identified candidate field, based on self-expressing confidence~\cite{xu2024sayself}. On the end of the prompt, a transaction instance is appended, serving as the specific input for the LLM to analyze and infer relevant information. An example of the prompt structure is provided in supplemental files.

For category $i$ and component $j \in \{A, T, D, C, Ts\}$, if the candidate identification results $QT_{cdd}^{i,j}$ candidate fields, the total number of possible quintuples is then reduced to $\mathcal{Y}$:

\begin{equation}
    \mathcal{Y}=\sum_{i=1}^{M} \left(\prod_{j} |QT_{cdd}^{i,j}|\right) \label{eq:possible_combination2}
\end{equation}



\stitle{Design Choices.}
The goal of this step is to rapidly identify a candidate set of potentially semantic fields for each element of the quintuple, drawn from a vast search space. While alternative methods exist, they present significant challenges:
(1) Rule-based or heuristic approaches (e.g., Connector~\cite{lin2025connector}) lack generalizability due to the diverse and evolving naming conventions in smart contracts. (2) Methods based purely on semantic similarity are often inefficient and prone to error, as incorrect fields can frequently exhibit higher similarity scores than correct ones. (3) Hybrid methods, which combine well-crafted heuristic rules with semantic similarity techniques, represent a viable alternative. 

We selected an LLM-based approach for its advanced semantic reasoning and generalization capabilities. LLMs can better interpret the contextual meaning of fields, overcoming the limitations of rigid rules and noisy similarity scores.
To validate this choice, we conduct a comprehensive comparison against heuristic ($B_{rule}$), similarity-based ($B_{sim}$), and hybrid ($B_{hybrid}$) baselines in \sectionname~\ref{subsubsec:baseline}. We also evaluate performance across different LLM backends in \sectionname~\ref{subsec:RQ3}.


\subsection{Validation}
\label{subsec:validation}

To identify and validate the correct field assignments for quintuples, \sysname construct an examiner algorithm $Ex$. As shown in \algorithmname~\ref{alg:quintuple_validation}, taking as input a set of candidate quintuples ($QT_{cdd}$) filtered by LLM, along with transactions of source chain ($TX_s$) and destination chain ($TX_d$), and a timewindow hyperparameter, $Ex$ refines these candidates to produce a validated quintuples ($QT_{corr}$). For each transaction on the source chain, $Ex$ proceeds in three phases. The first phase (\linename~\ref{algline:phase1_begin}-\ref{algline:phase1_end}) focuses on narrowing down candidate fields for the Amount ($A$) and Token ($T$) components. It begins by analyzing the asset flow within the source transaction to obtain a set of `(AmountValue, TokenType)' (\linename~\ref{algline:asset_flow}). Subsequently, for each combination of candidate Amount and Token fields retrieved from $QT_{cdd}$, their corresponding values are extracted (\linename~\ref{algline:ext_A_T}), and a matching operation is performed against the identified asset flows(\linename~\ref{algline:asset_flow_match}). Validated $\{A:F_A, T:F_T\}$ assignments are then added to $QT_{corr}$ (\linename~\ref{algline:add_A_T}).

The second phase aims to refine the candidate fields for Destination ($D$), Chain ($C$), and Timestamp ($Ts$) by identifying corresponding transactions on the destination chain. For each combination of candidate $D, C$, and $Ts$ fields and extracted values (\linename~\ref{algline:ext_D_C_Ts}), $Ex$ locate potential matching destination transactions within the specified timewindow, based on the extracted chain and timestamp values (\linename~\ref{algline:locate_dst_tx}). For each destination transaction $tx_d$, its candidate Destination field ($D$) is extracted (\linename~\ref{algline:ext_v_D_dst}), and a value comparison is performed between the destination address indicated by $V_D$ and $V_{D_{dst}}$ (\linename~\ref{algline:comp_v_vdst}). If a match is found, the corresponding validated $\{D:F_D, C:F_C, Ts:F_{Ts}\}$ assignment is recorded in $QT_{corr}$ (\linename~\ref{algline:add_D_C_Ts}). This two-stage validation process ensures that only field assignments consistent with observed asset flows and cross-chain transactional pairing are retained, providing a robust mechanism for identifying accurate quintuples.


The third phase is designed to refining the identified quintuples (A, T, D, C, Ts) to ensure their accuracy. For any component where multiple candidate fields have been associated (i.e., the set of potential fields for that component is not singular), two specific criteria are applied sequentially: consistency and uniqueness. 
The first criterion is consistency. The \varname{CheckConsistency} function verifies whether all candidate fields for a specific component (e.g., $F_A^1, F_A^2, \dots, F_A^n$) extract identical values across all transactions within the dataset $TX_s$ (\linename~\ref{algline:check_consistency}). If consistency is confirmed, it implies that all these candidate fields are functionally equivalent for that component, as they yield the same output. In such cases, any one of them can be validly selected. If the consistency criterion is not met (i.e., candidate fields for a component do not extract identical values), the process then evaluates the uniqueness of each individual candidate field. The \varname{isUnique} function assesses whether the value extracted by a candidate field $F_k^i$ from transactions in $TX_s$ is sufficiently variable to serve as a distinguishing identifier (\linename~\ref{algline:check_uniqueness}). Specifically, a field is deemed unsuitable if its extracted value remains constant across all transactions. A constant value prevents the field from differentiating one cross-chain transaction from another, thus failing its role as an identifier. Candidate fields identified as non-unique are subsequently removed from the set of potential fields(\linename~\ref{algline:remove_constant}). The details of the two functions (i.e., \varname{CheckConsistency} and \varname{isUnique}) are shown in \algorithmname~\ref{alg:helper_func}. Finally, $Ex$ returns the identified quintuple $QT_{corr}$. If $TX_s$ is is grouped into $M$ distinct categories, the total number of possible quintuples is reduced to $M$. 

While the algorithm, as presented, primarily details the process for identifying quintuple components originating from source chain transactions, its core logic is generalizable. To adapt $Ex$ for identifying quintuples on the destination chain, only a conceptual reversal of the 'source' and 'destination' chain roles is required. This effectively means interchanging the sets of transactions (e.g., treating $TX_d$ as the primary input for iteration instead of $TX_s$) and adjusting any chain-specific field retrieval or matching logic accordingly.

\begin{algorithm}[!t]
    \SetAlgoLined 
    \DontPrintSemicolon 
    \SetKwComment{tcp}{\text{//}}{\text{}} 
    \SetKwProg{Fn}{Function}{:}{}
    \KwIn{ Candidate quintuples $QT_{cdd}$, transactions on source chain $TX_s$ and destination chain $TX_d$, hyperparameter $timewindow$}
    \KwOut{the identified correct quintuple $QT_{corr}$}

    $QT_{corr} \leftarrow \varnothing$ \;
    \ForEach{$tx_s \in TX_s$}{ \label{algline:phase1_begin}
        \tcp{Phase1: filter base on asset flows}
        $flows \leftarrow \Call{AnalyzeAssetFlow}{tx_s}$\; \label{algline:asset_flow}
        $F_{A}^{all}, F_{T}^{all} \leftarrow \KwGet{QT_{corr}}{\text{`src'}, tx_s, (A,T)}$ \;
        \ForEach{$(F_{A}, F_{T}) \in (F_{A}^{all}, F_{T}^{all})$}{
            $V_{A}, V_{T} \leftarrow \Call{Extract}{(F_{A}, F_{T})}$\; \label{algline:ext_A_T}
            \If{$\Call{FlowMatch}{flows, (V_{A}, V_{T} )}$}{ \label{algline:asset_flow_match}
                $\KwAdd{QT_{corr}}{ \{ A:F_{A}, T:F_{T} \} }$ \; \label{algline:add_A_T} \label{algline:phase1_end}
            }
        }

        \tcp{Phase2:value matching}
        $F_{D}^{all}, F_{C}^{all}, F_{Ts}^{all} \leftarrow \KwGet{QT_{corr}}{\text{`src'}, tx_s,(D,C,Ts)}$ \;
        \ForEach{$ (F_{D}, F_{C}, F_{Ts}) \in (F_{D}^{all}, F_{C}^{all}, F_{Ts}^{all} )$}{
            $V_{D}, V_{C}, V_{Ts} \leftarrow \Call{Extract}{tx_s,F_{D}, F_{D}, F_{Ts}}$\; \label{algline:ext_D_C_Ts}
            \ForEach{$tx_d \in \Call{FindByCTs}{TX_d, V_{C}, V_{Ts}, timewindow}$}{ \label{algline:locate_dst_tx}
                $F_{D_{dst}} \leftarrow \KwGet{QT_{cdd}}{\text{`dst'}, tx_d, D} $ \;
                $V_{D_{dst}} \leftarrow \Call{Extract}{tx_d, F_{D_{dst}}}$ \; \label{algline:ext_v_D_dst}
                \If{$\Call{ValueMatch}{V_{D}, V_{D_{dst}}}$}{ \label{algline:comp_v_vdst}
                    $\KwAdd{QT_{corr}}{\{ D:F_{D}, C:F_{C},Ts:F_{Ts}\}}$\; \label{algline:add_D_C_Ts} 
                }
            }
        }
    }
    
    \tcp{Phase3: refine}
    \ForEach{$k \in \{A,T,D,C,Ts\} $}{
        $F_{k} \leftarrow \KwGet{QT_{corr}}{k} $ \;
        \If{ $ F_{k}.\text{length} > 1$ }{
            \If{ \Call{CheckConsistency}{$TX_s, F_{k}$} } { \label{algline:check_consistency}
                 \textbf{continue}\; 
            }
            \ForEach{$F_k^{i} \in F_k$ }{
                \If{ not \Call{isUnique}{$TX_s, F_k^{i}$} }{ \label{algline:check_uniqueness}
                    \KwRemove{$QT_{corr}$}{$F_k^{i}$}\; \label{algline:remove_constant}
                }
            }
        }
    }

    \Return $QT_{corr}$ \;

    \caption{Cross-Chain Quintuple Examination}
    \label{alg:quintuple_validation}

\end{algorithm}

\begin{algorithm}[thb]
    \SetAlgoLined 
    \DontPrintSemicolon 
    \SetKwComment{tcp}{\text{//}}{\text{}} 
    \SetKwProg{Fn}{Function}{:}{}

    \BlankLine
    \Fn{CheckConsistency($TX, F_k$)}{
        $F_{k}^{1},\dots, F_{k}^{n} \leftarrow F_{k} $ \;
        \ForEach{$tx \in TX $ }{
            $V_{k}^{1},\dots, V_{k}^{n} \leftarrow \Call{Extract}{tx, (F_{k}^{1},\dots, F_{k}^{n}) }$  \;
            \If{ not \Call{AllSame}{$V_{k}^{1},\dots, V_{k}^{n}$} }{
                \Return `false'\;
            }
        }
        \Return `true' \;
    }

    \BlankLine
    \Fn{isUnique($TX, F_k^{i}$)}{
        $ValueSet \leftarrow \varnothing $ \;
        \ForEach{$tx \in TX $ }{
            $V^{i} \leftarrow \Call{Extract}{tx, F_{k}^{i}}$\;
            $\KwAdd{ValueSet}{V^{i}}$ \;
        }
        \Return $\KwLength{ValueSet} != 1$ \;
    }

    \caption{Helper Functions of Examination}
    \label{alg:helper_func}
\end{algorithm}

\subsection{Pairing}
\label{subsec:pairing}


In this subsection, we describe how to establish pairing relationships between cross-chain transactions based on the quintuples validated in \sectionname~\ref{subsec:validation}. 

\stitle{Pairing Rules.}
For each transaction instance, \sysname first extracts five key pieces of information from the quintuple: destination address ($D$), token type ($T$), asset amount ($A$), chain identifier ($C$), and timestamp ($Ts$). 
Then, given any pair of transaction instances residing on different blockchains, \sysname attempts to match them according to the following rules (shown in \tablename~\ref{tab:matching_rules}): 
(1) One instance must represent the initiating transaction (sender), while the other represents the receiving transaction (receiver). 
(2)-(4) The destination address ($D$), token type ($T$), and amount of the two instances must correspond. The asset amount is also expected to be consistent, allowing for a discrepancy no greater than a defined $fee\_rate$. 
(5) The destination chain of one transaction must be the source chain of the other, and vice-versa. 
(6) The timestamps of the two transactions must fall within a reasonable time window ($timewindow$). 
According to our experiments on different parameters (\sectionname~\ref{subsec:hyper_para}), the hyper-parameters $timewindow$ and $fee\_rate$ are set to 2 hours and 20\%, respectively.

\begin{table}[!t]
    \centering
    \caption{Formal Pairing Rules for Cross-Chain Transactions Pairing.}
    \label{tab:matching_rules}
    \resizebox{\linewidth}{!}{
    \begin{tabular}{@{}l|@{ }r@{ }}
      \toprule
      \textbf{Condition} & \textbf{Formal Representation} \\
      \midrule
      (1) Role & $isSender(t_s) \land isReceiver(t_d)$ \\
      \midrule
      (2) Destination & $to(t_s) = to(t_d)$  \\
      \midrule
      (3) Token Type & $token(t_s) = token(t_d)$  \\
      \midrule
      (4) Amount & $ \frac{|amount(t_s)-amount(t_d)|}{amount(t_s)} \leq fee\_rate $ \\
      \midrule
      \multirow{2}{*}{(5) Chain} & $chain(t_d) = dstChain(t_s)$ \\ 
      & $\land chain(t_s) = srcChain(t_d)$ \\
      \midrule
      (6) Timestamp & $|timestamp(t_s)-timestmap(t_d)| \leq timewindow$ \\
      \bottomrule      
    \end{tabular}
    }
\end{table}


\stitle{Handling of Multiple Matches.} 
In cases where a single source transaction instance matches multiple destination transaction instances, \sysname selects the destination transaction instance with the earliest timestamp. This prioritizes the earliest chronologically potential match.
\section{Evaluation}
\label{sec:evaluation}






We designed 4 Research Questions (RQs) to comprehensively evaluate our \sysname:

\begin{rqitem}
  \item Is \sysname effective in terms of identifying cross-chain transaction pairs? How does it compare to other baseline techniques?
  \item How efficient is \sysname in terms of pruning searching space and reducing runtime? 
  \item Will the effectiveness of \sysname be seriously affected with different LLMs? 
  \item How does \sysname performs given different hyperparameters?
  \item What are the application scenarios for \sysname?
\end{rqitem}

\subsection{Experiment Setup}
\label{subsec:exp_setup}

We implement \sysname with Python and evaluate it using three popular LLMs: GPT-4o (version 2024-0806), Gemini-2.0-Flash, and Deepseek-R1. All experiments are conducted on an Ubuntu 24.04 server (Kernel version 6.8.0) equipped with dual Intel Xeon Gold 6252 processors (24 cores/48 threads per CPU), 256GB RAM, and a 22TB HDD.

\subsubsection{Dataset}
\label{subsubsec:dataset}


The construction of our dataset involved selecting both cross-chain bridges and the constituent blockchain networks. 
For cross-chain bridge selection, our primary criterion was the public accessibility of cross-chain transaction records. Accordingly, we selected DLN~\cite{DLN_bridge} and Stargate~\cite{Stargate}, identified among the top-ranked bridges (\tablename~\ref{table:bridges_rank}). In addition, we incorporated a publicly available dataset from prior research, Connector~\cite{lin2025connector}, encompassing Multichain~\cite{Multichain_bridge}, Celer Bridge~\cite{Celer_cBridge}, and PolyNetwork~\cite{Polygon_bridge} (abbreviated as Multi, Celer, Poly, respectively).
Regarding blockchain selection, we selected four of the most widely used EVM-compatible blockchains (Ethereum, Base, Optimism, and Arbitrum), based on the popularity rankings presented in \tablename~\ref{table:chains_rank}. These chains represent a significant portion of the cross-chain activity in the ecosystem.

Following the selection, we proceeded with data collecting. We manually identified the router contract addresses for DLN and Stargate through their official documentation. Subsequently, we use RPC endpoints to collect transactions related to cross-chain operations, along with their corresponding event logs. Transactions of the rest three bridges (Multi, Celer, Poly) are collected from the Connector dataset. This data collection process yielded a dataset comprising 503,627 transactions and 2,515,710 event logs associated with the five cross-chain bridges. 
\tablename~\ref{table:dataset_summary} provides key statistics of the dataset.

\begin{table}[!t]
    \centering
    \caption{Statistical Information of Our Dataset. The data of Multi, Celer and Poly bridge are from Connector~\cite{lin2025connector}, which includes 3 chains: Ethereum, Polygon, and BSC. We include data from Stargate and DLN from 4 chains: Ethereum, Base, Optimism, and Arbitrum.}
    \label{table:dataset_summary}
    \resizebox{\linewidth}{!}{
    \begin{tabular}{@{ }ll@{ }c@{ }c@{ }@{ }c@{ }}
        \toprule
        \textbf{Bridge}  & \textbf{Date Range} & \textbf{\#chains}  & \textbf{\#~txs} & \textbf{\#~logs} \\
        \midrule
            \textbf{Stargate} & 2024.03.01-2024.03.10   & 4 & 340,243         & 1,832,278 \\
            \textbf{DLN}      & 2024.03.01-2024.03.20 & 4 & 80,533          & 497,754\\
            \textbf{Multi}    & 2021.04.15-2022.04.11   & 3 & 20,980          & 81,034 \\
            \textbf{Celer}    & 2021.12.02-2023.05.03  & 3 & 15,788          & 33,856\\
            \textbf{Poly}     & 2021.02.04-2023.05.03   & 3 & 12,010          & 70,788\\ 
        \bottomrule
    \end{tabular}
    }
\end{table}

\subsubsection{Baselines}
\label{subsubsec:baseline}

\begin{figure}[!t]
  \centering
  \includegraphics[width=1.0\linewidth]{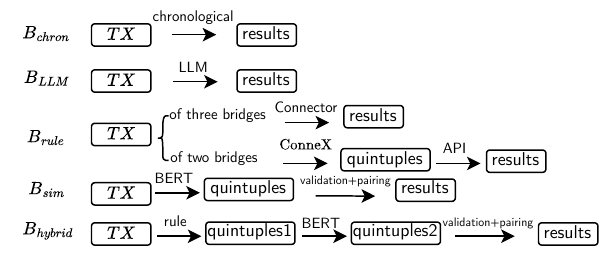}
  \caption{Illustration of Baselines.}
  \label{fig:baseline_illustration}
\end{figure}

As shown in \figurename~\ref{fig:baseline_illustration}, we constructed 5 baselines to enable comparison, including chronological ($B_{chron}$), LLM-only($B_{LLM}$), rule-based($B_{rule}$), similarity-based($B_{sim}$) and hybrid($B_{hybrid}$) methods:

\stitle{Baseline 1: Chronological Order ($B_{chron}$).}
This baseline examines whether cross-chain transaction matching can be effectively achieved through simple temporal ordering. 
To achieve better pairing, we first manually align the initial transaction on the source chain with its corresponding destination chain transaction. Subsequent transactions are then paired chronologically. 
This tests the hypothesis that simply connecting transaction instances on either side of a cross-chain bridge in chronological order can yield correct pairs.

\stitle{Baseline 2: LLM Only ($B_{LLM}$).}
This baseline aims to isolate the contribution of the LLM itself, demonstrating that the performance of \sysname is not solely attributable to the LLM's inherent reasoning capabilities. For each initiating transaction on a given source chain, the LLM is tasked with selecting the most probable matching destination transaction from the pool of candidate transactions on the destination chain.
To manage computational complexity and context length limitations inherent to LLMs, we constrained the search to destination transactions occurring within a time window of $timewindow$ = 2 hours after the initiation of the source transaction. This time window aligns with the hyperparameter defined in \sectionname~\ref{subsec:pairing}. To provide a fair comparison, both evaluations of $B_{LLM}$ and \sysname utilize the same underlying LLM (Gemini).

\stitle{Baseline 3: Rule-based by existing work ($B_{rule}$).}
This baseline incorporates methodology from the related work Connector~\cite{lin2025connector}. While Connector primarily focuses on identifying whether a transaction is part of any cross-chain activity, it provides transaction matching results for three specific bridges relevant to our study: Multi, Poly, and Celer. For these three bridges, we directly utilize the matching results reported by Connector in their evaluation. 
Connector designs manually defined rule-based methods for extracting transaction semantics from these three bridges, but did not consider Stargate and DLN, resulting in the ineffectiveness of analyzing the latter two bridges.
Therefore, a hybrid approach is necessary for these two bridges. Specifically, we leverage \sysname to extract the semantic information from Stargate and DLN transactions, which is then supplied as input to Connector's pairing mechanism, and subsequently utilizes the Etherscan API to query potential matches and generate pairing results.
This hybrid strategy allows for a comparison that leverages Connector's established Etherscan interaction logic while evaluating the effectiveness of \sysname's semantic extraction capabilities on bridges not natively supported by Connector's rule sets.

\stitle{Baseline4: Similarity Based Filtering ($B_{sim}$).}
To justify the employing of an LLM for semantic filtering, this baseline replaces the LLM's role (\sectionname~\ref{subsec:query_llm}) with a pre-trained embedding model. Specifically, it utilizes Sentence-BERT~\cite{reimers2019sentence} to select the top 5 semantically closest candidates for each element of the quintuple from a vast search space. The descriptions of the quintuple elements and each possible semantic candidate are first parsed as sentences and then encoded into embeddings. For examle, the correct semantics in \figurename~\ref{fig:motivation_example_C2} (marked in {\color{color_of_correct_answer}\charBlock}) will be parsed into a sentence `\texttt{log createOrder receiver Dst}'. Next, for each element of the quintuple, we use cosine similarity to find the 5 semantically closest candidates. The subsequent verification and generation steps then follow the design of \sysname (\sectionname~\ref{subsec:validation}-\ref{subsec:pairing}). 

\stitle{Baseline5: Heuristic Base + Similarity Based Filtering ($B_{hybrid}$).}
To evaluate the effectiveness of a hybrid approach combining heuristics with semantic similarity, we designed this baseline that first applies a heuristic filtering based on the data type of each semantic element to narrow down the candidate pool. Subsequently, from this pre-filtered set, it employs the same semantic similarity-based method as $B_{sim}$ to make the final selection.
Specifically, we define the following types:


to: string

token: string

amount: integer

chain: integer

timestamp: integer

\subsection{RQ1: Effectiveness}
\label{subsec:RQ1}



\begin{table}[!t]
	\centering
	\caption{Effectiveness of Each Cross-Chain Bridge.}
	\label{table:compare_baseline_bridges}
	\resizebox{\linewidth}{!}{
		\begin{tabular}{@{ }l@{ }@{ }l@{ }@{ }r@{ }@{ }r@{ }@{ }r@{ }@{ }r@{ }@{ }r@{ }@{ }r@{ }}
			\toprule
			\textbf{Bridge}                    & \textbf{Metrics} & \textbf{$B_{chron}$} & \textbf{$B_{LLM}^{\natural}$} & \textbf{$B_{rule}$}         & \textbf{$B_{sim}$} & \textbf{$B_{hybrid}$} & \textbf{\sysname}                    \\
			\midrule
			\multirow{3}{*}{\textbf{Stargate}} & precision        & 0.0005               & 0.0784                        & 0.8730$^{\flat}$            & 0.0000             & 0.0000                & 0.9972                               \\
			                                   & recall           & 0.0034               & 0.9620                        & 0.6709$^{\flat}$            & 0.0000             & 0.0000                & 0.9923                               \\
			                                   & F1 score         & 0.0008               & 0.1450                        & 0.7587$^{\flat}$            & 0.0000             & 0.0000                & \underline{\textbf{0.9948}}          \\
			\midrule
			\multirow{3}{*}{\textbf{DLN}}      & precision        & 0.0005               & 0.1251                        & 0.7697$^{\flat}$            & 0.9831             & 0.9834                & 0.9981                               \\
			                                   & recall           & 0.0029               & 0.9741                        & 0.0943$^{\flat}$            & 0.0877             & 0.0889                & 0.8653                               \\
			                                   & F1 score         & 0.0008               & 0.2218                        & 0.1680$^{\flat}$            & 0.1612             & 0.1631                & \underline{\textbf{\textbf{0.9270}}} \\
			\midrule
			\multirow{3}{*}{\textbf{Multi}}    & precision        & 0.0008               & 0.8176                        & 0.9940                      & 0.9986             & 0.9986               & 0.9986                               \\
			                                   & recall           & 0.0008               & 0.9849                        & 0.9856                      & 0.4079             & 0.4055                & 0.9714                               \\
			                                   & F1 score         & 0.0008               & 0.8935                        & \underline{\textbf{0.9898}} & 0.5792             & 0.5768                 & 0.9848                               \\
			\midrule
			\multirow{3}{*}{\textbf{Celer}}    & precision        & 0.0005               & 0.7725                        & 0.9937                      & 0.9991             & 0.9985                & 0.9984                               \\
			                                   & recall           & 0.0005               & 0.9778                        & 0.9937                      & 0.8380             & 0.9600                & 0.9682                               \\
			                                   & F1 score         & 0.0005               & 0.8631                        & 0.9766                      & 0.9115             & 0.9789                & \underline{\textbf{0.9831}}          \\
			\midrule
			\multirow{3}{*}{\textbf{Poly}}     & precision        & 0.0030               & 0.9722                        & 0.9921                      & 0.9850             &  0.9862                & 0.9862                               \\
			                                   & recall           & 0.0030               & 0.9877                        & 0.9324                      & 0.8761             &  0.9647               & 0.9754                               \\
			                                   & F1 score         & 0.0030               & 0.9799                        & 0.9613                      & 0.9274             &  0.9753               & \underline{\textbf{0.9807}}          \\
			\midrule
			\midrule
			\multirow{3}{*}{\textbf{Total}}    & precision        & 0.0010               & 0.5531                        & 0.9149                      & 0.7931             &  0.7933               & 0.9957                               \\
			                                   & recall           & 0.0021               & 0.9772                        & 0.6708                      & 0.4419             &  0.4838               & 0.9545                               \\
			                                   & F1 score         & 0.0014               & 0.7064                        & 0.7741                      & 0.5676             &  0.6010               & \underline{\textbf{0.9746}}          \\
			\bottomrule
		\end{tabular}
	}

	\caption*{\small $^{\flat}$: As Connector's rule-based approach for src transaction semantic extraction cannot apply to Stargate and DLN, we use \sysname's output for src semantic. \par $^{\natural}$: To save token, we sample 10\% results to examine.}
\end{table}
\begin{table}[!t]
	\centering
	\caption{Effectiveness (F1 Score) of Each Blockchain Pair. }
	\label{table:compare_baseline_chains}
	\resizebox{\linewidth}{!}{
	\begin{tabular}{@{ }l@{ }@{ }r@{ }@{ }r@{ }@{ }r@{ }@{ }r@{ }@{ }r@{ }@{ }r@{ }}
		\toprule
		\textbf{src $\rightarrow$ dst$^{\flat}$} & \textbf{$B_{chron}$} & \textbf{$B_{LLM}$} & \textbf{$B_{rule}$} & \textbf{$B_{sim}$} & \textbf{$B_{hybrid}$} & \textbf{\sysname}     \\
		\toprule
		\textbf{E $\rightarrow$ A}               & 0.0016               & 0.1041             & 0.5235              & 0.0000             & 0.0322                & \underline{\textbf{0.9120}} \\
		\textbf{E $\rightarrow$ B}               & 0.0015               & 0.2432             & 0.1815              & 0.0000             & 0.0000                & \underline{\textbf{0.9517}} \\
		\textbf{E $\rightarrow$ O}               & 0.0023               & 0.0259             & 0.3459              & 0.0000             & 0.0000                & \underline{\textbf{0.8626}} \\
		\textbf{E $\rightarrow$ BSC}             & 0.0010               & 0.9488             & 0.9229              & 0.8157             & 0.8455                & \underline{\textbf{0.9843}} \\
		\textbf{E $\rightarrow$ P}               & 0.0027               & 0.4779             & 0.8798              & 0.6430             & 0.7800                & \underline{\textbf{0.9765}} \\
		\textbf{A $\rightarrow$ E}               & 0.0019               & 0.0514             & 0.4627              & 0.6692             & 0.6692                & \underline{\textbf{0.9069}} \\
		\textbf{A $\rightarrow$ B}               & 0.0004               & 0.1751             & 0.6590              & 0.0000             & 0.0000                & \underline{\textbf{0.9465}} \\
		\textbf{A $\rightarrow$ O}               & 0.0007               & 0.2619             & 0.4776              & 0.0000             & 0.0000                & \underline{\textbf{0.9591}} \\
		\textbf{B $\rightarrow$ E}               & 0.0024               & 0.0789             & 0.1053              & 0.9002             & 0.9055                & \underline{\textbf{0.9459}} \\
		\textbf{B $\rightarrow$ A}               & 0.0005               & 0.2074             & 0.1355              & 0.0000             & 0.0000                & \underline{\textbf{0.9821}} \\
		\textbf{B $\rightarrow$ O}               & 0.0008               & 0.1705             & 0.1082              & 0.0000             & 0.0000                & \underline{\textbf{0.9701}} \\
		\textbf{O $\rightarrow$ E}               & 0.0025               & 0.0284             & 0.3345              & 0.7910             & 0.7910                & \underline{\textbf{0.8848}} \\
		\textbf{O $\rightarrow$ A}               & 0.0007               & 0.4300             & 0.6052              & 0.0000             & 0.0000                & \underline{\textbf{0.9648}} \\
		\textbf{O $\rightarrow$ B}               & 0.0005               & 0.2701             & 0.6136              & 0.0000             & 0.0000                & \underline{\textbf{0.9467}} \\
		\bottomrule
	\end{tabular}
	}
	\caption*{\small $^{\flat}$: E=Ethereum, A=Arbitrum, P=Polygon, B=Base, O=Optimism.}
\end{table}

To evaluate its effectiveness, \sysname was benchmarked against several baseline methods using our collected dataset. The results, detailed in \tablename~\ref{table:compare_baseline_bridges} (per-bridge) and \tablename~\ref{table:compare_baseline_chains} (per-chain pair), show that \sysname significantly outperforms all baselines.

\sysname achieved a consistently higher F1-score (0.9746), exceeding the baselines by an average of 97.3\%, 26.8\%, 20.05\% ,40.7\% and 37.3\%, respectively.
This improvement was particularly significant for the DLN and Stargate bridges, attributable to two key factors.
First, \sysname effectively addresses the inherent challenges in these cases, such as a larger candidate answer pool and potential misleading information (see \sectionname~\ref{sec:motivating_example}), whereas baseline methods fail to handle them robustly. Second, unlike the baselines (especially baseline $B_{rule}$), which are limited to native assets and ERC20 token transfers, \sysname supports non-ERC20 tokens (e.g., Stargate’s sgETH), enabling more accurate pairing relationships for DLN and Stargate.
The superior performance in these challenging scenarios demonstrates the generality of \sysname in handling diverse cross-chain transfer types.


\stitle{Answer to RQ1:} \sysname, which attains an average F1 score of 0.9746, is effective in identifying cross-chain transaction pairs, and exceeding the baselines at least 20.05\%.

\subsection{RQ2: Ablation Study}
\label{subsec:abalation}

This section evaluates the efficiency of \sysname by analyzing its search space reduction and runtime performance.

\subsubsection{Searching Space}
\label{subsubsec:ablation_search_space}


\tablename~\ref{table:abalation_res} illustrates the reduction in the number of possible quintuple selections after each processing stage of \sysname. As detailed in \sectionname~\ref{sec:design}, Step \ding{173} involves categorizing transaction instances by grouping them into $M$ distinct categories, which results in a possible combination numbers $\mathcal{X}$ (\eqname~\ref{eq:possible_combination1}). Subsequently, Step \ding{174} employs an LLM to refine the selection within each category, which results in a possible combination $\mathcal{Y}$ (\eqname~\ref{eq:possible_combination2}). Finally, Step \ding{175} further prunes these possibilities, leading a total number of possible quintuples reducing to $M$ (yielding one quintuple per category). 
The result demonstrates the significant efficiency of our filtering methodology in drastically reducing a vast semantic search space (over 1\textbf{e}10) to a manageable and effective set of selections (two-digit range).

\subsubsection{Runtime Performance}
\label{subsubsec:ablation_time}


\tablename~\ref{table:time_performance} reports the runtime performance of \sysname. We first detail the execution time of Step \ding{174} (i.e., querying the LLM). Furthermore, we established a simple baseline to illustrate the runtime difference with and without the validation step (Step \ding{175} in \sectionname~\ref{subsec:validation}). When the validation step is omitted, \sysname resorts to a brute-force approach, attempting all permutations and combinations of candidates identified by the LLM. Each such candidate necessitates a comprehensive cross-chain asset flow analysis to verify whether its corresponding semantic meaning truly materialized on the other side of the bridge.
As reported in \tablename~\ref{table:time_performance}, \sysname is capable of processing all bridge providers within several hours. When averaged per transaction across each bridge, the processing time is less than 1 second. Compared to the baseline without the validation step, \sysname achieves a significant speedup, reaching up to 9 times faster performance (e.g., when processing Celer). 

\stitle{Answer to RQ2: }
\sysname is efficient in terms of pruning searching space and reducing runtime. Notably, it prunes a vast semantic search space (over 1\textbf{e}10) into a manageable set of selections (less than 100), and reaches up to 9x faster runtime performance.

\begin{table}[!t]
    \centering
    \caption{Number of Possible Quintuple Choices after Each System Step.}
    \label{table:abalation_res}
    
    \resizebox{0.9\linewidth}{!}{
    
    \begin{tabular}{@{ }l|@{ }r@{ }r@{ }r@{ }|@{ }r@{ }r@{ }r@{ }}
        \toprule
        \multicolumn{1}{c}{\multirow{2}{*}{}} & \multicolumn{3}{c}{\textbf{src}}        & \multicolumn{3}{c}{\textbf{dst}}      \\
        \midrule
        \multicolumn{1}{c}{}                  & \textbf{Step\ding{173}}       & \textbf{Step\ding{174}} & \textbf{Step\ding{175}} & \textbf{Step\ding{173}}     & \textbf{Step\ding{174}} & \textbf{Step\ding{175}} \\
        \midrule
        \textbf{Stargate}                              & 2\textbf{e}7     & 3\textbf{e}3   & 21     & 4\textbf{e}9 & 6\textbf{e}4  & 344    \\
        \textbf{DLN}                                   & 1\textbf{e}12 & 2\textbf{e}5 & 144    & 2\textbf{e}9 & 7\textbf{e}4  & 81     \\
        \textbf{Multi}                                 & 1\textbf{e}5       & 1\textbf{e}3   & 7      & 1\textbf{e}5     & 6\textbf{e}2    & 4      \\
        \textbf{Celer}                                 & 8\textbf{e}4        & 1\textbf{e}3   & 9      & 5\textbf{e}5     & 1\textbf{e}3   & 8      \\
        \textbf{Poly}                                  & 4\textbf{e}6      & 2\textbf{e}3   & 6      & 9\textbf{e}4      & 549    & 6      \\
        \bottomrule
    \end{tabular}
    }
\end{table}

\begin{table}[!t]
    \centering
    \caption{Runtime Performance (minutes) of \sysname. }
    \label{table:time_performance}
    \begin{tabular}{lrrrr}
        \toprule
        \textbf{Bridge}   & \textbf{Step\ding{174}} & \textbf{w/o \ding{175}} & \textbf{w/ \ding{175}}  &\textbf{Per Tx} \\
        \midrule
        \textbf{Stargate} & 160 & 679 & 568                 & 0.0016                     \\
        \textbf{DLN}       & 339& 1,774 & 421                  & 0.0052                     \\
        \textbf{Multi}    & 5& 235 & 30                    & 0.0014                     \\
        \textbf{Celer}    & 13& 346 & 38                     & 0.0024                     \\
        \textbf{Poly}      & 5& 219 & 42                    & 0.0034                     \\
        \bottomrule            
    \end{tabular}
\end{table}

\subsection{RQ3: Performance on different LLMs}
\label{subsec:RQ3}
\begin{table}[!t]
	\centering
	\caption{Effectiveness of Different LLMs.}
	\label{table:different_LLM_v}

	\begin{tabular}{@{ }l@{ }@{ }l@{ }@{ }r@{ }@{ }r@{ }@{ }r@{ }}

		\toprule
		\textbf{Bridge}                    & \textbf{Metrics} & \textbf{GPT-4o}             & \textbf{Gemini} & \textbf{Deepseek}        \\
		\midrule
		\multirow{3}{*}{\textbf{Stargate}} & precision        & 0.9971                      & 0.9972          & 0.9965                      \\
		                                   & recall           & 0.9926                      & 0.9923          & 0.7891                      \\
		                                   & F1 score         & \underline{\textbf{0.9949}} & 0.9948          & 0.8807                      \\
		\midrule
		\multirow{3}{*}{\textbf{DLN}}      & precision        & 0.9983                      & 0.9981          & 0.9990                      \\
		                                   & recall           & 0.9312                      & 0.8653          & 0.9157                      \\
		                                   & F1 score         & \underline{\textbf{0.9636}} & 0.9270          & 0.9555                      \\
		\midrule
		\multirow{3}{*}{\textbf{Multi}}    & precision        & 0.9985                      & 0.9986          & 0.9986                      \\
		                                   & recall           & 0.9837                      & 0.9714          & 0.9800                      \\
		                                   & F1 score         & \underline{\textbf{0.9911}} & 0.9848          & 0.9892                      \\
		\midrule
		\multirow{3}{*}{\textbf{Celer}}    & precision        & 0.9987                      & 0.9984          & 0.9986                      \\
		                                   & recall           & 0.9719                      & 0.9682          & 0.9772                      \\
		                                   & F1 score         & 0.9851                      & 0.9831          & \underline{\textbf{0.9877}} \\
		\midrule
		\multirow{3}{*}{\textbf{Poly}}     & precision        & 0.9862                      & 0.9862          & 0.9862                      \\
		                                   & recall           & 0.9755                      & 0.9754          & 0.9755                      \\
		                                   & F1 score         & \underline{\textbf{0.9808}} & 0.9807          & 0.9808                      \\
		\midrule
		\midrule
		\multirow{3}{*}{\textbf{Total}}    & precision        & 0.9958                      & 0.9957          & 0.9958                      \\
		                                   & recall           & 0.9709                      & 0.9545          & 0.9275                      \\
		                                   & F1 score         & \underline{\textbf{0.9832}} & 0.9746          & 0.9604                      \\
		\bottomrule
	\end{tabular}
\end{table}


To evaluate \sysname's robustness across varying LLM architectures and capabilities, we conducted experiments using GPT-4o, Gemini, and Deepseek. The results, summarized in \tablename~\ref{table:different_LLM_v}, reveal that while GPT-4o achieved the highest performance, the F1-score difference between the models remained below 3 percentage points. This relatively small variance suggests that \sysname's performance is not critically tied to a specific LLM. Importantly, even in the worst-case scenario (i.e., using Deepseek), \sysname still outperforms the baselines by over 18\%. This indicates that \sysname maintains a significant level of effectiveness regardless of the LLM backend, ensuring practical deployability and reducing reliance on a single LLM provider. 

\stitle{Answer to RQ3: } \sysname maintains high F1 score when employing different LLM backends, which suggests the robustness of design logic. 



\subsection{RQ4: Impact of Hyperparameters}
\label{subsec:hyper_para}

\begin{table}[th]
	\centering
	\caption{Performance (F1 score) of Different Hyper-Parameter Configurations.}
	\label{table:hyper_para}
	\begin{tabular}{@{ }c|c@{ }@{ }c@{ }@{ }c@{ }@{ }c@{ }@{ }c@{ }c@{ }}
		\toprule
		\diagbox[dir=NW]{$fr$}{ $tw$} & \textbf{10} & \textbf{60} & \textbf{600} & \textbf{3600} & \textbf{7200} & \textbf{10800} \\
		\midrule
		\textbf{0.01} & 0.0000 & 0.0000 & 0.8484 & 0.9154 & 0.8666 & 0.9177 \\
		\textbf{0.05} & 0.0000 & 0.0000 & 0.8487 & 0.9157 & 0.9628 & 0.9637 \\
		\textbf{0.1}  & 0.0000 & 0.0000 & 0.8565 & 0.9683 & 0.9690 & 0.9245 \\
		\textbf{0.15} & 0.0000 & 0.0000 & 0.9047 & 0.9726 & 0.9255 & 0.9709 \\
		\textbf{0.2}  & 0.0000 & 0.0000 & 0.8595 & 0.9257 & 0.9746 & 0.9274 \\
		\bottomrule
	\end{tabular}
	\caption*{$fr$ = $fee\_rate$, $tw$ = $timewindow$}
\end{table}


To investigate the impact of hyperparameter selection ($timewindow$ and $fee\_rate$) in \sysname, we conducted experiments across 25 different configurations, varying the $timewindow$ (10, 60, 600, 3,600, 7,200 and 10,800 seconds) and the $fee\_rate$ (0.01, 0.05, 0.1, 0.15 and 0.2). These ranges were selected based on common time delay and fee structures mentioned in the documents of studied bridges. For each configuration, we recorded the average F1 score across all evaluated bridges. 
The experimental results are presented in \tablename~\ref{table:hyper_para}. It was observed that when the hyperparameter $timewindow$ was set to a very small value ($< 60$s), valid experimental results could not be obtained. This is attributed to the fact that such a limited time window significantly impedes the effective identification of potentially corresponding transactions on another chain.

\stitle{Answer to RQ4:} As both $fee\_rate$ and $timewindow$ increase, \sysname consistently achieves a higher F1 score. Specifically, with relatively relaxed parameter settings ($fee\_rate \ge 0.15$ and $timewindow \ge 3600$s), the F1 score stabilizes at a satisfactory level (exceeding 0.96).


\subsection{RQ5: Application of \sysname}
\label{subsec:application}


\begin{figure}[!t]
    \centering
    \includegraphics[width=1.0\linewidth]{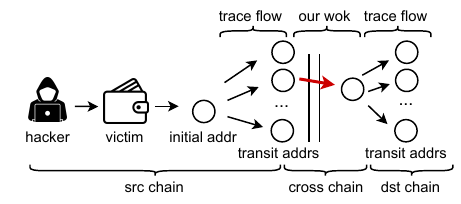}
    \caption{An Illustration of the Downstream Task: Tracing Money Flow across Blockchains.}
    \label{fig:downstream_task}
\end{figure}


This section presents \sysname's application to cross-chain money laundering analysis, an area where conventional fund tracing tools often fail due to their limitations in cross-chain transaction pairing.

\figurename~\ref{fig:downstream_task} shows a typical money laundering scheme, where hackers exploit vulnerabilities to steal funds and consolidate them at a hacker-controlled address. Hackers then obfuscate the origin of these funds by dispersing them through multiple intermediary addresses.
A key tactic involves using cross-chain bridges to transfer stolen assets from a source chain to a destination chain. 
Current money laundering analysis techniques~\cite{wu2023toward} often struggle to track funds across these bridges due to the lack of accurate cross-chain relationship pairing. 
This limitation restricts investigations to isolated blockchains (e.g., Ethereum), forcing analysts to abandon tracking once funds enter a cross-chain bridge.
Consequently, a critical blind spot emerges, giving hackers an advantage in evading detection.

\sysname addresses this challenge by enabling analysts and researchers to accurately pair transactions across different blockchains, thereby extending the scope of money laundering investigations beyond the boundaries of a single chain. To illustrate the practical benefits of \sysname in real-world scenarios, we present two case studies of actual hacks where cross-chain bridges were likely used to launder stolen funds.

\subsubsection{Case 1: Bybit Hack}
\label{subsubsec:app_example1}

Reports \cite{bybit2025bybitSecurity} indicate that hackers stole hundreds of thousands of ETH from Bybit's cold wallet and subsequently distributed them to numerous intermediary addresses.
We initially identified these intermediary addresses, which were already flagged by existing methods and easily discoverable through basic fund flow tracking using Etherscan. Subsequent analysis of the fund flows originating from these addresses revealed a critical cross-chain transaction initiated from one address via the DLN bridge\footnote{\url{https://etherscan.io/tx/0x538a296a4995dd3aea4a4a2f7db423d795458e409a64b2776575326558126a2b}}. 
We use \sysname to analyze this critical transaction, leveraging the quintuple identified for the DLN bridge in \sectionname~\ref{subsec:RQ1}:

to: trx.\_orderCreation.receiverDst

token: trx.\_orderCreation.giveTokenAddress

amount: trx.\_orderCreation.giveAmount

chain: trx.\_orderCreation.takeChainId

timestamp: trx.timestamp

By extracting the corresponding values from the transaction, we determined the target chain to be Solana, with a receiving address 0xc4..f0\footnote{Solana uses base58-encoded address foramt, therefore the final address is EFmqz8PTTShNsEsErMUFt9ZZx8CTZHz4orUhdz8Bdq2P}. The transferred asset was 1,086,388 USDC at timestamp 1740207179. 
By searching the transaction records of the DLN bridge on Solana, we found that the hacker's intermediary address received 1,085,510 USDC through DLN approximately three minutes later\footnote{\url{https://solscan.io/tx/5daJgqjkb4bd7KUhQFpZwEzRw6gCZVtT7y4rUXpMf87J44vvQL7DL2QTmep7nXyvUm7vqgWyzCqGKspinP9Q6CQD}}. 
This spotted money laundering activity is also confirmed by other security researchers and Solana officials, demonstrating the effectiveness of \sysname in tracing illicit funds across blockchains.

\subsubsection{Case 2: Upbit Hack}
\label{subsubsec:app_example2}

Previous work \cite{wu2023toward} analyzed and tracked the fund flows associated with the Upbit hack, identifying approximately 20,000 suspicious addresses. Further analysis of the transactions involving these addresses revealed that the hackers initiated a cross-chain transfer\footnote{\url{https://etherscan.io/tx/0x47f034003e7809f6701ff1c0020873a211739b5b1fdcd710e1db5780241b5bdb}} to Optimism using the official Optimism Bridge. Specifically, a transfer of 0.1 ETH was made to an address with the same address on Optimism. Notably, shortly after receiving the funds on Optimism, the hacker immediately used the cross-chain bridge again to transfer the same amount of ETH back to the same address on Ethereum\footnote{\url{https://optimistic.etherscan.io/tx/0x1c1a4b6d228e47c79324ba2fb352d90fad0a2fb6dfed123103175337851dd982}}. This rapid back-and-forth transfer between chains, involving the same address and amount, is strongly indicative of funds laundering, where the goal is to obfuscate the origin and destination of the illicit funds. Without \sysname, such circular cross-chain transactions could be misinterpreted as legitimate activity, leading to inaccurate assessments of fund flow and potentially hindering effective money laundering investigations. 





\stitle{Answer to RQ5:} \sysname is helpful for the downstream task of cross-chain money laundering analysis with real-world cases. Notably, in the Bybit Hack, \sysname successfully identifies a transit address receiving \$1 million USDCs.




\section{Discussion}
\label{sec:discussion}

\stitle{FPs \& FNs.}
Despite its demonstrated effectiveness(\sectionname~\ref{subsec:RQ1}), \sysname is subject to limitations that produce both false positives (FPs) and false negatives (FNs). 
For \textbf{FPs}, they are caused by inaccuracies in the quintuple of transactions and by overly permissive hyperparameter settings, which can lead to the incorrect matching of unrelated transactions. For \textbf{FNs}, they stem from two main issues: (1)Overly restrictive hyperparameters can cause \sysname to miss valid transaction pairs. This highlights a fundamental trade-off between minimizing FPs and FNs.(2) Discrepancies between the token types used in a transaction pair (e.g., ETH on a source chain and WETH on a destination chain) make it difficult to determine asset equivalence, leading to missed matches. To mitigate the token heterogeneity issue, we implemented a value normalization method. This approach uses price oracles to convert transaction amounts to a standardized unit for comparison. Currently, this normalization is limited to high-volume assets such as stablecoins and Wrapped ETH. Extending this conversion mechanism to support arbitrary tokens remains a key direction for future work.

\stitle{Bridge Scope.}
The scope of this study is restricted to the cross-chain bridge scope defined in \sectionname~\ref{sec:problem_definition}. This focus facilitates a simplified analysis of core bridge mechanisms. However, the proposed method is adaptable. For instance, it could accommodate L1 to L2 bridges (e.g., Arbitrum Bridge~\cite{Arbitrum_bridge}) by relaxing the `chain' parameter definition, or bridges with one-to-many transaction patterns (e.g., Meson \cite{Meson}) by adjusting the transaction matching heuristics.

\stitle{ABI Dependency.}
Following prior work~\cite{lin2025connector}, we assume the availability of router contract ABIs. This assumption is practical, as studies show approximately 80\% of bridge contracts are open-source~\cite{lin2025connector}, with operators often verifying their source code on public platforms like Etherscan to foster trust. For close-source router contracts, it is still possible to infer partial ABI information using existing decompilation tools (e.g., gigahorse~\cite{grech2019gigahorse}, Elipmoc~\cite{grech2022elipmoc}).

\section{Conclusion}
\label{sec:conclusion}

This paper presented \sysname, a novel system that addresses a critical security challenge in the multi-chain Web3 ecosystem: the difficulty of tracking assets across blockchains due to implicit transaction pairings. \sysname automates the process of accurately pairing source and destination transactions.
\sysname employs LLMs for pruning and a specialized examiner module for validation via key-value pairs. Experimental results on real-world datasets demonstrate its high accuracy and efficiency, significantly outperforming baseline methods. Furthermore, its practical utility was validated by successfully identifying intermediary addresses in actual cross-chain money laundering incidents.


\ifthenelse{\TemplateUsenix}{
    \cleardoublepage
    \appendix
    \section*{Ethical Considerations}
\label{sec:ethical_consideration}


We conduct our ethical analysis using a stakeholder-based framework, guided by the principles of Beneficence, Respect for Persons, Justice, and Respect for Law and Public Interest, as outlined in The Menlo Report.

\stitle{Security, Privacy, and Dual-Use Nature.}
The primary ethical tension arises from a conflict between enhancing security and preserving privacy. \sysname promotes the principle of \textit{Beneficence} by providing a mechanism to trace illicit funds from security breaches, as demonstrated in our case study (\sectionname~\ref{subsec:application}). However, this capability concurrently diminishes the practical pseudonymity of blockchain transactions, which raises concerns under the principle of \textit{Respect for Persons} regarding individual financial privacy.

We acknowledge the dual-use nature of this technology. We contend that similar analytical capabilities are likely already developed and used privately by sophisticated malicious actors. Withholding this research would thus perpetuate an information asymmetry that disadvantages the broader defensive community. By publishing our work, we aim to level this asymmetry, arming defenders and improving the collective security of the ecosystem.

\stitle{Mitigation Strategies.}
To mitigate potential harms, we have taken several deliberate steps:
\begin{itemize}[leftmargin=*]
    \item Our research is strictly confined to publicly available and verifiable blockchain data. No private APIs, servers, or personally identifiable information were accessed.
    \item We are publishing our methodology, system design, and evaluation, not a user-friendly, push-button tool. This raises the barrier for misuse while enabling legitimate researchers and developers to understand, verify, and build upon our work for defensive purposes.
    \item Our real-world application focuses on a widely publicized, major hacking incident. This demonstrates utility against acknowledged criminal activity without scrutinizing the transactions of normal, unsuspecting users.
    \item Our analysis stops at linking pseudonymous on-chain addresses. We have made no attempt to deanonymize these addresses to real-world identities.
\end{itemize}

\stitle{Decision to Publish.}
Finally, our decision to proceed with and publish this research is based on the conviction that its benefits to the security and integrity of the multi-chain ecosystem substantially outweigh the mitigated risks. This decision aligns with the principle of \textit{Respect for Law and Public Interest}, as it contributes to a community-wide defense against illicit activities in the Web3 space..

    \cleardoublepage
    
    \section*{Open Science}
\label{sec:open_science}

We make our artifacts available at \artifactUrl, including collected data, codes and results.  
    \cleardoublepage
}{}

\ifthenelse{\TemplateACM}{


\bibliographystyle{ACM-Reference-Format}
\bibliography{reference}

}{\ifthenelse{\TemplateIEEE}{


\bibliographystyle{IEEEtranS}
\bibliography{reference}

}{\ifthenelse{\TemplateUsenix}{

    \bibliographystyle{plain}
    \bibliography{reference}

}{} 
}}

\ifthenelse{\TemplateACM\or\TemplateUsenix}{
\appendix
}{  

\appendices
}

\end{document}